\documentclass[12pt]{article}
\IfFileExists{hyperref.sty}{\usepackage{hyperref}}{}

\newtheorem{Lemma}{Lemma}
\newtheorem{Proposition}[Lemma]{Proposition}
\newtheorem{Definition}[Lemma]{Definition}

\def\qed{\leavevmode \hfill\hbox to.77778em{\hfil\vrule
         \vbox to.675em{\hrule width.6em\vfil\hrule}\vrule\hfil}}

\def\semi{\mathop{>\!\!\!\triangleleft}}

\makeatletter
\def\section{\@startsection{section}{1}{\z@}{-3.25ex 
       plus -1ex minus -.2ex}{1.5ex plus .2ex}{\normalfont\bfseries}}
\def\subsection{\@startsection{subsection}{2}{\z@}{-3.25ex 
        plus -1ex minus -.2ex}{1.5ex plus .2ex}{\normalfont\itshape}} 
\@addtoreset{equation}{section}

\def\ps@mathphys{\addtolength{\headheight}{5pt}
            \addtolength{\topmargin}{-15pt}
            \addtolength{\headsep}{15pt}
    \def\@oddhead{\hfil\begin{tabular}{r}
          {\tt math-ph/9904009} \\ CPT-99/P.3809
          \end{tabular}}
          \let\@evenhead\@oddhead
          \def\@oddfoot{\hfil\thepage\hfil}\let\@evenfoot\@oddfoot}

\makeatother

\font\twelvebb=msbm10 scaled 1200
\newfam\bbfam
\textfont\bbfam=\twelvebb
\def\bb{\fam\bbfam}

\begin{document}

\thispagestyle{mathphys}

\begin{center}
{\baselineskip 20pt
{\large\bf On the Connes--Moscovici Hopf algebra associated to
the diffeomorphisms of a manifold}
\vskip 3ex

{\sc Raimar Wulkenhaar}
}
\vskip 1ex 
 
{\it Centre de Physique Th\'eorique\\
CNRS-Luminy, Case 907\\
13288 Marseille Cedex 9, France}

\end{center}
\vskip 3ex

\begin{abstract}
For our own education, we reconstruct the Hopf algebra of Connes and
Moscovici obtained by the action of vector fields on a crossed product
of functions by diffeomorphisms. We extend the realization of that
Hopf algebra in terms of rooted trees as given by Connes and Kreimer
from dimension one to arbitrary dimension of the manifold. In
principle there is no modification, but in higher dimension one has to
be careful with the order of cuts. The order problem leads us to
speculate that in quantum field theory the sum of Feynman graphs which
corresponds to an element of the Connes--Moscovici Hopf algebra could
have a larger symmetry than the individual graphs.
\\[1ex]
\tabcolsep 0pt 
\begin{tabular}{lll}
1991 MSC:~~{} & 16W30~{} & Coalgebras, bialgebras, Hopf algebras \\
& 57R25 & Vector fields, frame fields \\
& 57R50 & Diffeomorphisms \\
& 81T15 & Perturbative methods of renormalization 
\end{tabular}
\end{abstract}

\section{Introduction}

Recently two useful Hopf algebras were discovered -- by Alain Connes
and Henri Moscovici in noncommutative geometry \cite{cm2} and by Dirk
Kreimer in quantum field theory \cite{k1}. Connes and Kreimer showed
that both Hopf algebras are intimately related \cite{ck}, via the
language of rooted trees. Recently it was pointed out \cite{b} that
the same algebra of rooted trees appears in numerical analysis. We
refer to \cite{ck2} for a review of all these ideas.

For a physicist, the Hopf algebra ${\cal H}_K$ of Kreimer \cite{k1} is
not difficult to understand. The idea is to look at the divergence
structure of Feynman graphs. There is a natural splitting of a Feynman 
graph $\gamma$ with non-overlapping subdivergences into two, given by
a product of selected subdivergences $\gamma_1\cdots \gamma_n$ and the
graph $\gamma\setminus(\gamma_1\cdots \gamma_n)$ left over from
$\gamma$ by shrinking all $\gamma_i$, $i=1,\dots,n$, to a point. This
is a standard operation in renormalization theory. Summing over all
possibilities gives a coproduct operation on the algebra of
polynomials in Feynman graphs. The unique antipode reproduces
precisely the combinatorics of renormalization, i.e.\ it produces the
local counterterms to make the integral corresponding to the divergent
Feynman graph finite.

The aim of this note is to review in some detail the construction of
the Hopf algebra found by Connes and Moscovici, in order to ease its
access for physicists interested in the Hopf algebra ${\cal H}_K$ of
renormalization. The construction requires only some basic knowledge
of classical differential geometry, which can be found in many books
on this topic, for instance in \cite{cd}. More precisely, one needs
the vertical and horizontal vector fields $Y$ and $X$ on the frame
bundle over an oriented manifold and their transformation behavior
under diffeomorphisms, as well as some familiarity with the
push-forward and pull-back operations. These requisites are derived in
section \ref{general}. New is the application of these vector fields
to the crossed product, see section \ref{cross}, which defines the
coproduct of $X,Y$ and leads to an additional operator $\delta$ on the
crossed product. The operators $X,Y,\delta$ generate a Lie algebra.
Its enveloping algebra ${\cal H}$ is a bialgebra with the coproduct
obtained before, and there exists an antipode making it to a Hopf
algebra, see section \ref{Lie}.

The final section is devoted to the transformation of the commutative
Hopf subalgebra ${\cal H}_{CM}$ of ${\cal H}$ into the language of
rooted trees so that we can compare it with ${\cal H}_K$. We
generalize the rooted trees given in \cite{ck} from dimension 1 to
arbitrary dimension of the manifold. This generalization is quite
obvious, but it has several consequences which are not visible in
dimension 1. An element of ${\cal H}_{CM}$ is a sum of decorated
planar rooted trees. The root is decorated by three spacetime indices
necessary to describe parallel transport whereas the other vertices
are decorated by a single spacetime index. This is closer to quantum
field theory, where the decoration is given by primitive Feynman
graphs without subdivergences. Interestingly, elements of ${\cal
H}_{CM}$ are invariant under permutations of the decorations, whereas
the individual trees representing Feynman graphs are not. This raises
the question whether the sum of Feynman graphs which corresponds to an
element of ${\cal H}_{CM}$ has a meaning in QFT.

\section{The geometry of the frame bundle}
\label{general}

In this section we are going to derive in some detail the following
well-known results on the principal fibre bundle $F^+$ of oriented
frames on an $n$-dimensional manifold $M$:

\begin{Proposition}
Let $\{x^\mu\}_{\mu=1\dots,n}$ be the coordinates of $x\in M$ within a
local chart of $M$ and $\{y^\mu_i\}_{\mu,i = 1,\dots n}$ be the
coordinates of $n$ linearly independent vectors of the tangent space
$T_xM$ with respect to the basis $\partial_\mu$. On $F^+$ there exist
the following geometrical objects, written in terms of the 
local coordinates $(x^\mu,y^\mu_i)$ of $p \in F^+$:
\begin{itemize}
\item[(1)] an ${\bb R}^n$-valued (soldering) $1$-form $\alpha$ with
$\alpha^i = (y^{-1})^i_\mu d x^\mu~,$

\item[(2)] a $gl(n)$-valued  (connection) $1$-form $\omega$ with 
$\omega^i_j = (y^{-1})^i_\mu ( d y^\mu_j + \Gamma^\mu_{\alpha\beta}\,
y^\alpha_j dx^\beta)$, where $\Gamma^\mu_{\alpha\beta}$ depends only
on $x^\nu~,$

\item[(3)] $n^2$ vertical vector fields $Y^i_j = y^\mu_j \partial^i_\mu~,$

\item[(4)] $n$ horizontal (with respect to $\omega$) vector fields 
$X_i = y^\mu_i ( \partial_\mu - 
\Gamma^\nu_{\alpha\mu} y^\alpha_j \partial^j_\nu)~.$
\end{itemize}
A local diffeomorphism $\psi$ of $M$ has a lift $\tilde{\psi}:
(x^\mu,y^\mu_i) \mapsto (\psi(x)^\mu,\partial_\nu \psi(x)^\mu
y^\nu_i)$ to the frame bundle and induces the following
transformations of the previous geometrical objects:
\begin{itemize}
\item[(1)] $(\tilde{\psi}^* \alpha)\big|_p = \alpha\big|_p~.$
\item[(2)] $(\tilde{\psi}^* \omega)\big|_p = (y^{-1})^i_\mu 
( d y^\mu_j + \tilde{\Gamma}^\mu_{\alpha\beta}\, y^\alpha_j dx^\beta)$ is
again a connection form, with 
\\[0.5ex]
$\tilde{\Gamma}^\mu_{\alpha\beta}\big|_x 
= ((\partial\psi(x))^{-1})^\mu_\gamma \,
\Gamma^\gamma_{\delta\epsilon}\big|_{\psi(x)} \,
\partial_\alpha\psi(x)^\delta  
\partial_\beta \psi(x)^\epsilon + ((\partial\psi(x))^{-1})^\mu_\gamma \,
\partial_\beta \partial_\alpha \psi(x)^\gamma ~,$

\item[(3)] $(\tilde{\psi}_* Y^j_i)\big|_p = Y^j_i\big|_p~,$

\item[(4)] $(\tilde{\psi}^{-1}_* X_i)\big|_p = y^\mu_i 
( \partial_\mu - \tilde{\Gamma}^\nu_{\alpha\mu} y^\alpha_j 
\partial^j_\nu)$ is horizontal to $\tilde{\psi}^* \omega~.$
\end{itemize}
\label{prpgen}
\end{Proposition}
The reader familiar with these notations can pass immediately to section
\ref{cross} on page~\pageref{cross}.

\subsection{Frame bundle}

Let $M$ be an $n$-dimensional smooth and oriented manifold. We are
going to consider the frame bundle $F^+$ over $M$ defined as
follows. Let $T_xM$ be the tangent space at a given point $x \in
M$. It is an $n$-dimensional vector space containing the tangent
vectors at $x$ of curves in $M$ through $x$. A base in $T_xM$ is given
by the $n$ tangent vectors $\partial_\mu:=\frac{\partial}{\partial
x^\mu}$ of the coordinate lines in $M$. If $x$ has
(in a given chart of its neighbourhood) the coordinates
$\{x^\mu\} \equiv(x^1,\dots,x^n)$, we compute the tangent vector to a
curve $C (t) = \{x^\mu(t)\}$: 
\begin{equation}
\frac{d\phi(C(t))}{dt}\Big|_{t=0} 
=\frac{d x^\mu}{dt}\Big|_{t=0} \; \frac{\partial}{\partial x^\mu} 
\phi \Big|_x~,
\end{equation}
where $\phi:M \to {\bb R}$ (or ${\bb C}$) is an arbitrary function on
$M$.  According to Einstein's sum convention summation over pairs of
identical upper and lower indices is self-understood.

An arbitrary vector $Y_j \in T_xM$ can be decomposed with respect to
that basis, $Y_j = y^\mu_j \partial_\mu$. A frame at $x$ is now a set of $n$
linearly independent vectors $Y_j \in T_xM$, $j=1,\dots,n$,
parameterized by their coordinates $y^\mu_j$, where both $\mu$ and $j$
run from $1$ to $n$. Linear independence is equivalent to $\det y \neq
0$, and oriented frames have the same sign of $\det y$.

The (oriented) frame bundle $F^+$ is now given by the base space $M$
with the set of smooth (positively oriented) frames attached to
each point $x \in M$. A point in $F^+$ is thus (locally) given by the
collection
\[
(x,\{Y_j\})=(x^\mu, y_j^\mu)_{\mu,j=1,\dots,n}~, \qquad \det y > 0~,
\]
where $x^\mu$ are the coordinates of $x$ and the $y^\mu_j \in Gl^+(n)$ 
parameterize an oriented frame $\{Y_j\}_{j=1,\dots,n}$ at $x$. Here, 
$Gl^+(n)$ is the group of $n \times n$ matrices with positive
determinant. 

In the overlap of two charts $U_1,U_2$, a point $x \in U_1 \cap U_2 
\subset M$ will have coordinates $x^\mu$ in $U_1$ and
$x^\nu{}'$ in $U_2$. The tangent vector $Y_i$ at a curve in $U_1 \cap
U_2$ through $x$ is given 
\begin{eqnarray*}
\mbox{in $U_1$ by}\quad{} &
Y_i &= \frac{df(x^\mu(t))}{dt}\Big|_{t=0} = \frac{dx^\mu(t)}{dt}
\Big|_{t=0} \; \partial_\mu f~,
\\
\mbox{in $U_2$ by}\quad{} &
Y_i &= \frac{df(x^\nu{}'(x^\mu(t)))}{dt}\Big|_{t=0} 
= \frac{dx^\mu(t)}{dt}\Big|_{t=0} \;
\frac{\partial x^\nu{}'}{\partial x^\mu} \;\partial_\nu' f~,
\end{eqnarray*}
where $f$ is an arbitrary function on $F^+$. Hence, the coordinates
$(x^\mu,y^\mu_j) \in U_1 \times Gl^+(n)$ and $(x^\nu{}',y^\nu_j{}')
\in U_2 \times Gl^+(n)$ label the same point in $F^+$ iff
$x^\mu,x^\nu{}'$ are the coordinates in $U_1,U_2$ of $x \in U_1 \cap
U_2$ and $y^\nu_j{}' = (\partial x^\nu{}'/\partial x^\mu) y^\mu_j$.

There is a natural action of $Gl^+(n)$ on a frame $\{Y_j\}$ at $x$: The
matrix $g^i_j \in Gl^+(n)$ maps the vector $Y_i \in T_xM$ into the new
vector $Y_i g^i_j=:Y'_j \in T_xM$, or~-- in coordinates~-- $y^\mu_i$
into $y^\mu_i g^i_j$. This $Gl^+(n)$-action extends naturally to the
frame bundle, making $F^+$ to a $Gl^+(n)$-principal fibre bundle:
\begin{equation}
g:\,(x^\mu,y^\mu_i) \mapsto (x^\mu,y^\mu_i g^i_j)~.
\label{right}
\end{equation}

The above action can be regarded as generated by a vector field
according to the following construction. Let $gl(n)$ be the Lie
algebra of $Gl^+(n)$. The exponential mapping assigns to $A \in gl(n)$
a curve $\exp(tA)$ in $Gl^+(n)$, which by (\ref{right}) induces a
field of curves through every point of $F^+$. This field of curves
provides us with a field of tangent vectors
\[
\frac{df(x^\mu,y^\mu_i [\exp (tA)]^i_j}{dt} \Big|_{t=0} 
= \frac{\partial f}{\partial (y^\mu_k \delta^k_j)} \;
\frac{d (y^\mu_i [\exp (tA)]^i_j)}{dt} \Big|_{t=0} 
=  A^i_j y^\mu_i \frac{\partial}{\partial y^\mu_j} f ~,
\]
where $f$ is a function on $F^+$. Hence, each such vector field
associated to $A \in gl(n)$ is generated by the following (vertical)
vector fields
\begin{equation}
Y_i^j = y^\mu_i \frac{\partial}{\partial y^\mu_j} \equiv y^\mu_i 
\partial^j_\mu~.
\label{Y}
\end{equation}
The vector field $A^\# = A^i_j Y^j_i$ associated to $A \in gl(n)$ is
called the fundamental vector field corresponding to $A$.

A somewhat tricky construction provides us with an ${\bb R}^n$-valued
$1$-form $\alpha$ on $F^+$, sometimes called soldering form or
canonical $1$-form. A point $p =(x,\{y^\mu_j\}) \in F^+$ assigns to a
vector $\tilde{V} \in T_xM$ a vector $\Phi_p(\tilde{V}) \in {\bb R}^n$
by decomposing $\tilde{V}$ with respect to the basis $Y_j=y^\mu_j
\partial_\mu$. Thus, $[\Phi_p(\tilde{V})]^j Y_j = \tilde{V}$. Now, the
${\bb R}^n$-valued $1$-form $\alpha$ is defined by
\begin{equation}
\alpha(V) \big|_p = \Phi_p (\pi_* V)~, \qquad V \in T_p F^+~.
\label{defalpha}
\end{equation}
By $\big|_p$ we denote the value of a differential form or a vector
field at the point $p \in F^+$. In (\ref{defalpha}), $\pi_*$ is the
differential of the vertical projection $\pi(x,\{y^\mu_j\}) = x$ which
projects the vector $V=V^\mu \partial_\mu + V^\mu_j \partial^j_\mu \in
T_p F^+$ into the vector $\pi_* V =V^\mu \partial_\mu \in
T_{\pi(p)}M$. In this notation we have $\pi_* V = V^\nu (y^{-1})^j_\nu
Y_j$, using the obvious definition $y^\mu_i(y^{-1})^i_\nu =
\delta^\mu_\nu$. This gives $[\Phi_p (\pi_* V)]^j = V^\nu
(y^{-1})^j_\nu$. On the other hand, decomposing $\alpha^i=\alpha^i_\mu
dx^\mu + \alpha^{ik}_\mu dy^\mu_k$ and using the definition
\[
dy^\mu_i(\partial^j_\nu) = \delta^j_i \delta^\mu_\nu~,\quad 
dx^\mu(\partial_\nu) = \delta^\mu_\nu~,\quad 
dy^\mu_i(\partial_\nu)=0~,\quad 
dx^\mu(\partial^j_\nu) =0
\]
of the pairing between covectors and vectors, we have 
$\alpha^j(V) = \alpha^j_\mu V^\mu + \alpha^{ji}_\mu V^\mu_i$, giving 
\begin{equation}
\alpha^j = (y^{-1})^j_\mu \,dx^\mu~.
\label{alpha}
\end{equation}
The definition (\ref{defalpha}), although involving local coordinates
in the construction, is independent of the choice of charts. Indeed,
if $p \in F^+$ has the coordinates $(x^\mu,y^\mu_j)$ and $(x^\nu{}', 
y^\nu_j{}')$ in two charts $U_1 \times Gl^+(n)$ and $U_2 \times
Gl^+(n)$ of $F^+$, with $y^\nu_j{}'=(\partial x^\nu{}'/\partial
x^\mu) y^\mu_j$, then $\pi_* V = V^\mu \partial_\mu \in T_{\pi(p)}
U_1$ and $\pi_* V = V^\nu{}' \partial_{\nu'} \in T_{\pi(p)}
U_2$, with $ V^\nu{}' = (\partial x^\nu{}'/\partial x^\mu)
V^\mu$. This means that $V^\mu (y^{-1})^j_\mu 
=  V^\nu{}' (y^{-1})^j_{\nu'} \in {\bb R}$ give the same value for 
$[\Phi_p (\pi_* V)]^j$.

\subsection{Connection}

A connection is the splitting of the tangent space $T_pF^+$ at $p \in
F^+$ into a direct sum of a vertical space $V_pF^+$ (generated by
$Y^i_j=y^\mu_j \partial^i_\mu$) and a horizontal space $H_pF^+$ such
that $H_{pg}F^+=R_{g*} H_pF^+$. In the last equation, $pg \in F^+$ is
the point obtained by the action (\ref{right}) of $g \in Gl^+(n)$ and
$R_{g*}$ is the induced push-forward of a vector in $T_pF^+$ to a
vector in $T_{pg}F^+$. If $V \in T_pF^+$ is the tangent vector of a
curve $p(t)$ in $F^+$ through $p$, then the push-forward $R_{g*}V$ is the
tangent vector of the curve $p(t)g$ through $pg$. Explicitly, 
let $f$ be a function on $F^+$ and $V=V_j^\mu \partial^j_\mu+ V^\mu
\partial_\mu \in T_pF^+$ be tangent to the curve
$C=(x^\mu(t),y^\mu_j(t))$ at $p$, i.e.
\begin{equation}
Vf=(df(p)/dt)=\big((dx^\mu/dt)\partial_\mu 
+ (dy^\mu_j/dt)\partial^j_\mu\big)f~.
\label{Vf}
\end{equation}
Then, the push-forward is given by 
\begin{equation}
(R_{g*}V)f = \frac{df(p(t)\,g)}{dt} = \frac{dx^\mu(t)}{dt}\partial_\mu f 
+ \frac{d (y^\mu_j(t)\, g^j_i)}{dt} 
\frac{\partial f}{\partial \hat{y}^\mu_i}~,\qquad 
\hat{y}^\mu_i := y^\mu_k g^k_i~.
\label{RV}
\end{equation}
Thus, we obtain 
\begin{equation}
R_{g*}V= V_j^\mu g^j_i \,\widehat{\partial^i_\mu} + V^\mu \partial_\mu~,  
\qquad \widehat{\partial^i_\mu} := \partial /\partial \hat{y}^\mu_i~.
\label{RgV}
\end{equation}

In practice the connection is most conveniently characterized by the
connection form $\omega$, a $gl(n)$-valued differential $1$-form with
the following properties: For a given matrix $A \in gl(n)$ let $A^\#=
A^i_j Y^j_i$ be the corresponding fundamental vector field. Then
$\omega$ is defined by 
\begin{equation}
\omega(A^\#) = A ~,\qquad
\omega\big|_{pg}(R_{g*} V) = g^{-1} \Big(\omega\big|_p(V) \Big)g~,
\label{omega}
\end{equation}
for $V \in T_pF^+$ and $g \in Gl^+(n)$. At the point
$p=(x^\mu,y^\mu_i)$, the components $\omega^i_j$ of the connection
form will have the decomposition
\[
\omega^i_j = W^{ik}_{j \mu} dy^\mu_k + W^{i}_{j\mu} dx^\mu ~,
\]
for certain functions $W$. From (\ref{Y}) we get immediately 
\[
A^i_j \equiv \omega^i_j(A^\#) = W^{ik}_{j\mu} y^\mu_l A^l_k~,
\]
which gives $W^{ik}_{j\mu} = \delta^k_j (y^{-1})^i_\mu$. This
suggests the following ansatz
\begin{equation}
\omega^i_j = (y^{-1})^i_\mu ( d y^\mu_j + \Gamma^\mu_{\nu\alpha} y^\nu_j
d x^\alpha)~,
\label{om}
\end{equation}
where $\Gamma^\mu_{\nu\alpha}$ is a so far undetermined function of
$x$ and $y$. Using (\ref{RgV}) we write down
\begin{eqnarray*}
\omega^i_j\big|_{pg}(R_{g*}V) 
&=& (g^{-1})_k^i(y^{-1})^k_\mu \big( d \hat{y}^\mu_j 
+ \Gamma^\mu_{\nu\alpha}\big|_{pg}\, y^\nu_k g^k_j 
d x^\alpha\big) \big( V_m^\nu g^m_l \widehat{\partial_\nu^l}
+ V^\mu \partial_\mu \big) 
\\
&=& (g^{-1})_k^i(y^{-1})^k_\mu V_m^\mu g^m_j 
+ (g^{-1})_k^i(y^{-1})^k_\mu \Gamma^\mu_{\nu\alpha}\big|_{pg} \,
V^\alpha y^\nu_k g^k_j ~.
\end{eqnarray*}
On the other hand, 
\begin{eqnarray*}
\big[g^{-1}\big(\omega\big|_{p}(V)\big)g\big]^i_j = 
(g^{-1})_k^i(y^{-1})^k_\mu V_m^\mu g^m_j 
+ (g^{-1})_k^i(y^{-1})^k_\mu \Gamma^\mu_{\nu\alpha}\big|_p \,
V^\alpha y^\nu_k g^k_j ~.
\end{eqnarray*}
Thus, the condition (\ref{omega}) tells us that
$\Gamma^\mu_{\nu\alpha}\big|_p=\Gamma^\mu_{\nu\alpha}\big|_{pg}$,
{\em which means that $\Gamma^\mu_{\nu\alpha}$ depends only on the
base point $x$.}

Now, the horizontal vector fields $X_i$ associated to the connection
are defined as the kernel of $\omega$ and the dual of $\alpha$,
\begin{equation}
\omega^i_j (X_k)=0~,\qquad \alpha^j(X_i)=\delta^j_i~.
\end{equation}
These equations are easy to solve:
\begin{equation}
X_i = y^\mu_i (\partial_\mu 
- \Gamma^\nu_{\alpha\mu} y^\alpha_j \partial^j_\nu)~.
\label{X}
\end{equation}

The torsion form $\Theta$ on $F^+$ is an ${\bb R}^n$-valued
differential $2$-form defined as the covariant derivative of $\alpha$,
\begin{equation}
\Theta^i = d \alpha^i + \omega^i_j \wedge \alpha^j~.
\end{equation}
Using (\ref{alpha}) and (\ref{om}) we compute 
\begin{eqnarray*}
\Theta^i &=& - (y^{-1})^i_\nu (y^{-1})^j_\mu \,d y^\nu_j \wedge dx^\mu 
+ (y^{-1})^i_\mu ( d y^\mu_j + \Gamma^\mu_{\nu\alpha} y^\nu_j
d x^\alpha) \wedge (y^{-1})^j_\beta dx^\beta \\
&=& (y^{-1})^i_\mu  \Gamma^\mu_{\nu\alpha} d x^\alpha \wedge dx^\nu ~.
\end{eqnarray*}
The torsion vanishes iff the connection coefficients are symmetric, 
$\Gamma^\mu_{\nu\alpha} =\Gamma^\mu_{\alpha\nu}$.

\subsection{Diffeomorphisms}

Let $\psi$ be a local (orientation preserving) diffeomorphism of
$M$. By push-forward it maps a frame $\{Y_j\}$ at $x \in M$ into the
frame $\{\psi_* Y_j\}$ at $\psi(x) \in M$. If $Y$ is the tangent
vector at $x$ of a curve $C=\{x^\mu(t)\}$ through $x$, then $\psi_*
Y$ is the tangent vector at $\psi(x)$ of the curve
$\psi(C)=\{\psi(x(t))^\nu\}$. We evaluate both vectors on a function
$\phi$ on $M$: 
\begin{eqnarray*}
Y \phi &=& \frac{d}{dt} \phi(x^\mu(t)) \Big|_{t=0} 
= \frac{\partial \phi(x^\mu)}{\partial x^\mu} \:
\frac{dx^\mu(t)}{dt} \Big|_{t=0} 
= (dx^\mu(t)/dt)\big|_{t=0} \:\partial_\mu \phi~,
\\
(\psi_* Y) \phi &=& \frac{d}{dt} \phi (\psi(x(t))^\nu) \Big|_{t=0} 
= \frac{\partial \phi(\tilde{x}^\nu)}{\partial \tilde{x}^\mu} \:
\frac{\partial \psi(x)^\mu}{\partial x^\nu} \:
\frac{dx^\nu(t)}{dt} \Big|_{t=0} \\
&=& \partial_\nu \psi(x)^\mu\:
(dx^\nu(t)/dt)\big|_{t=0} \: \widetilde{\partial_\mu} \phi~,
\end{eqnarray*}
with $\tilde{x}^\mu = \psi(x)^\mu$ and $\widetilde{\partial_\mu}
=\partial/\partial \tilde{x}^\mu$.  Recall that $\partial_\mu$ and
$\widetilde{\partial_\mu}$ are the bases of vector fields in $T_xM$
and $T_{\psi(x)}M$, respectively. Hence, if $y^\mu_j$ are the
coordinates of $Y_j \in T_xM$, then $\psi_* Y_j \in T_{\psi(x)}M$ has
the coordinates $\partial_\nu \psi(x)^\mu \, y^\nu_j$, with respect to
these bases. To summarize, the action $\tilde{\psi}$ on $F^+$ of a
diffeomorphism $\psi$ of $M$ is given by
\begin{equation}
\tilde{\psi}: (\{x^\mu\}, \{y^\mu_i\}) \mapsto 
(\{\tilde{x}^\mu := \psi(x)^\mu\}, \{\tilde{y}^\mu_i 
:= \partial_\nu \psi(x)^\mu \, y^\nu_i\}) ~.
\label{psi}
\end{equation}
Note that the (right) action (\ref{right}) of
$Gl^+(n)$ on $F^+$ and the (left) action (\ref{psi}) on $F^+$ of a
diffeomorphism of $M$ commute with each other.

We consider now the effect of a diffeomorphism $\psi$ of $M$ on the
horizontal vector fields $X_i$. We use the following 
\begin{Lemma} 
The pull-back $\tilde{\psi}^* \omega$ of the connection form via the
action (\ref{psi}) of the induced diffeomorphism $\tilde{\psi}$ of
$F^+$ is again a connection form.
\end{Lemma}
{\it Proof.} We start from (\ref{Vf}) and compute 
\begin{eqnarray}
(\tilde{\psi}_* V) f &=& \frac{d}{dt} f\Big(\tilde{x}^\mu(x^\nu(t)), 
\tilde{y}^\mu_i(y^\nu_j(t),x^\nu(t))\Big) 
\nonumber
\\
&=& \frac{\partial f}{\partial \tilde{x}^\mu} 
\frac{\partial \psi(x)^\mu}{\partial x^\alpha} \frac{dx^\alpha}{dt} 
+ \frac{\partial f}{\partial \tilde{y}^\mu_i} \Big(  
\frac{\partial^2 \psi(x)^\mu}{\partial x^\alpha \partial x^\beta}
y_i^\beta \frac{dx^\alpha}{dt}
+ \partial_\alpha \psi(x)^\mu \frac{d y^\alpha_i}{dt} \Big)
\nonumber
\\
&=& \partial_\alpha \psi(x)^\mu V^\alpha \widetilde{\partial_\mu} f 
+ \big( \partial_\alpha \partial_\beta \psi(x)^\mu y_i^\beta V^\alpha 
+ \partial_\alpha \psi(x)^\mu V^\alpha_i \big)
\widetilde{\partial^i_\mu} f~,
\label{psiV}
\end{eqnarray}
where $\widetilde{\partial^i_\mu}= \frac{\partial}{\partial
\tilde{y}^\mu_i}$. For $V=A^\#$ we have $V^\mu=0$
and $V^\mu_i=A^j_i y^\mu_j$, see (\ref{Y}). This means 
\begin{equation}
\big(\tilde{\psi}_* A^\# \big)\big|_{\tilde{\psi}(p)}= 
A^j_i \partial_\alpha \psi(x)^\mu y^\alpha_j
\widetilde{\partial^i_\mu} = 
A^j_i \widetilde{y^\alpha_j} \widetilde{\partial^i_\mu} = A^\# 
\big|_{\tilde{\psi}(p)}~.
\label{Ainv}
\end{equation}
{\it The fundamental vector field $A^\#$ is invariant under
diffeomorphisms.} This gives 
\[
(\tilde{\psi}^*\omega)(A^\#)\big|_p=\omega(\tilde{\psi}_*A^\#)
\big|_{\tilde{\psi}(p)} =\omega(A^\#) \big|_{\tilde{\psi}(p)} = A~.
\]

The second identity to prove is
\begin{eqnarray*}
&&(\tilde{\psi}^* \omega)\big|_{pg}(R_{g*}V)\big|_{pg} = 
g^{-1} \Big((\tilde{\psi}^* \omega)(V)\big|_p \Big) g \quad
\\
&& \Rightarrow \quad 
\omega \big|_{\tilde{\psi}(pg)} \big( \psi_* \big((R_{g*}V)\big|_{pg}
\big) \big|_{\tilde{\psi}(pg)} \big)
= g^{-1} \Big( \omega\big|_{\tilde{\psi}(p)} (\tilde{\psi}_* V)
\big|_{\tilde{\psi}(p)} \Big) g ~.
\end{eqnarray*}
According to (\ref{RV}) we replace $V^\mu_i$ by $V^\mu_j g^j_i$ and
$y^\mu_i$ by $y^\mu_j g^j_i$ and insert this into (\ref{psiV}):
\[
\tilde{\psi}_* \big(R_{g*}V \big)\big|_{\tilde{\psi}(pg)} 
=  \partial_\alpha \psi(x)^\mu V^\alpha \widetilde{\partial_\mu} 
+ \big( \partial_\alpha \partial_\beta \psi(x)^\mu y_j^\beta V^\alpha 
+ \partial_\alpha \psi(x)^\mu V^\alpha_j \big) g^j_i 
\widetilde{\widehat{\partial^i_\mu}} ~,
\]
where $\widetilde{\widehat{\partial^i_\mu}}=
\partial/\partial \widetilde{\widehat{y^\mu_i}}$ and
$\widetilde{\widehat{y^\mu_i}}= \partial_\alpha \psi(x)^\mu 
y^\alpha_j g^j_i$. We must now evaluate 
\[
\omega^a_b \big|_{\tilde{\psi}(pg)} = (g^{-1})^a_c (y^{-1})^c_\gamma 
((\partial\psi(x))^{-1})^\gamma_\delta \Big( d
\widetilde{\widehat{y^\delta_b}} 
+ \Gamma^\delta_{\epsilon\zeta}\big|_{\psi(x)} \,
\partial_\eta\psi(x)^\epsilon y^\eta_d g^d_b \, d \widetilde{x^\zeta} 
\Big)
\] 
on the above vector:
\begin{eqnarray}
&& \omega^a_b \big|_{\tilde{\psi}(pg)}
\big(\tilde{\psi}_* \big(R_{g*}V \big)\big|_{\tilde{\psi}(pg)} \big) 
\nonumber \\
&&= (g^{-1})^a_c \Big\{(y^{-1})^c_\gamma 
((\partial\psi(x))^{-1})^\gamma_\delta \Big( \big(
\partial_\alpha \partial_\beta \psi(x)^\delta y_d^\beta V^\alpha 
+ \partial_\alpha \psi(x)^\delta V^\alpha_d \big) 
\nonumber \\ 
&&{} \hspace*{5cm}
+  \Gamma^\delta_{\epsilon\zeta}\big|_{\psi(x)} \,
\partial_\eta\psi(x)^\epsilon y^\eta_d \, \partial_\nu \psi(x)^\zeta
V^\nu \Big) \Big\} g^d_b ~. 
\label{qed}
\end{eqnarray}
Taking $g=e$ (identity matrix), it is obvious that the term in braces
$\{~\}$ equals $\omega^c_d (\tilde{\psi}_* V)
\big|_{\tilde{\psi}(p)}$, which finishes the proof of the
Lemma. \hfill \qed
\vskip 2ex

We can now rewrite the term in braces in (\ref{qed}) in a slightly
different way:
{\arraycolsep 0pt
\begin{eqnarray*}
&& \omega^c_d (\tilde{\psi}_* V)\big|_{\tilde{\psi}(p)} 
= \big(\tilde{\psi}^* \omega^c_d\big) (V) \big|_p 
\\ &&=  (y^{-1})^c_\gamma V^\gamma_d + (y^{-1})^c_\gamma 
((\partial\psi(x))^{-1})^\gamma_\delta \Big( 
\partial_\alpha \partial_\beta \psi(x)^\delta 
+  \Gamma^\delta_{\epsilon\zeta}\big|_{\psi(x)} \,
\partial_\beta\psi(x)^\epsilon  
\partial_\alpha \psi(x)^\zeta \Big) y_d^\beta V^\alpha 
\\
&&= (y^{-1})^c_\gamma ( d y^\gamma_d +
\tilde{\Gamma}^\gamma_{\beta\alpha}\big|_x\, y^\beta_d
d x^\alpha)(V^\nu_k \partial^k_\nu + V^\nu \partial_\nu)~,
\end{eqnarray*}}%
where $\tilde{\Gamma}^\gamma_{\beta\alpha}$ are the connection
coefficients of the connection $\tilde{\psi}^* \omega$.
This provides us with the following transformation law for the
connection coefficients:
\begin{equation}
\tilde{\Gamma}^\gamma_{\beta\alpha}\big|_x = 
((\partial\psi(x))^{-1})^\gamma_\delta \,
\Gamma^\delta_{\epsilon\zeta}\big|_{\psi(x)} \,
\partial_\beta\psi(x)^\epsilon  
\partial_\alpha \psi(x)^\zeta + ((\partial\psi(x))^{-1})^\gamma_\delta \,
\partial_\alpha \partial_\beta \psi(x)^\delta ~.
\label{Gamma}
\end{equation}

Now there is an immediate question to ask: Which are the horizontal
vector fields $\tilde{X}_i$ to the new connection form 
$\tilde{\psi}^* \omega$? We have
\[
0 = (\tilde{\psi}^* \omega)\big|_p (\tilde{X}_i \big|_p) = 
\omega \big|_{\tilde{\psi}(p)} (\tilde{\psi}_* \tilde{X}_i)
\big|_{\psi(p)} ~,
\]
which tells us 
\begin{equation}
\tilde{X}_i \big|_p = \tilde{\psi}^{-1}_*
\big(X_i\big|_{\tilde{\psi}(p)} \big) =  y^\mu_i (\partial_\mu 
- \tilde{\Gamma}^\nu_{\alpha\mu} \big|_x\,y^\alpha_j \partial^j_\nu)~.
\label{tX}
\end{equation}

The action (\ref{psi}) preserves the ${\bb R}^n$-valued $1$-form
$\alpha$ given in (\ref{defalpha}) and (\ref{alpha}). Indeed, for $V =
V^\mu \partial_\mu + V^\mu_i \partial^i_\mu \in T_pF^+$ we compute
using (\ref{psiV})
\begin{eqnarray*}
(\tilde{\psi}^* \alpha^j)\big|_p (V) &=& 
\alpha\big|_{\tilde{\psi}(p)} (\tilde{\psi}_*V) 
\\ 
&=& (\tilde{y}^{-1})^j_\mu d \tilde{x}^\mu \Big( 
\partial_\alpha \psi(x)^\mu V^\alpha \widetilde{\partial_\mu} 
+ \big( \partial_\alpha \partial_\beta \psi(x)^\mu y_i^\beta V^\alpha 
+ \partial_\alpha \psi(x)^\mu V^\alpha_i \big)
\widetilde{\partial^i_\mu} \Big) 
\\
&=& (\tilde{y}^{-1})^j_\mu \partial_\alpha \psi(x)^\mu V^\alpha 
= (y^{-1})^j_\alpha V^\alpha \equiv \alpha^j\big|_p (V)~.
\end{eqnarray*}

\section{Crossed product}
\label{cross}

The properties listed in Proposition \ref{prpgen} and derived
throughout section \ref{general} are the basis for the construction of
the Hopf algebra of Connes and Moscovici \cite{cm2}. The idea is
to apply the vertical and horizontal vector fields $Y^j_i$ and $X_i$
to a crossed product ${\cal A}$ defined below and to derive their
coproduct from
\begin{equation}
X_i (ab) = \Delta(X_i)\,(a \otimes b)~,\qquad 
Y^j_i (ab) = \Delta(Y^j_i)\,(a \otimes b)~,\qquad a,b \in {\cal A}~.
\label{gendelta}
\end{equation}
We refer to \cite{m} for an introduction to Hopf algebras and related
topics. 

Let $\Gamma$ be the pseudogroup of local (orientation preserving)
diffeomorphisms of $M$. We consider the crossed product of the algebra
$C^\infty_c(F^+)$ of smooth functions with compact support on the
frame bundle $F^+$ by the action of $\Gamma$,
\begin{equation}
{\cal A} = C^\infty_c(F^+)  \semi \Gamma~.
\end{equation}
As a set, ${\cal A}$ can be regarded as the tensor product of
 $C^\infty_c(F^+)$ with $\Gamma$. It is generated by the monomials
\begin{equation}
f U^*_\psi~,\qquad f \in C^\infty_c({\rm Dom}(\tilde{\psi}))~,\quad 
\psi \in \Gamma~,
\end{equation}
where $\tilde{\psi}$ is the diffeomorphism of $F^+$ induced by $\psi
\in \Gamma$ according to (\ref{psi}). As an algebra, the
multiplication rule in ${\cal A}$ is defined by
\begin{equation}
f_1 U^*_{\psi_1} \; f_2 U^*_{\psi_2} := f_1 (f_2 \circ \tilde{\psi}_1) 
U^*_{\psi_2 \psi_1} ~.
\label{prod}
\end{equation}
In this formula, the function $f_1 (f_2 \circ \tilde{\psi}_1) \in
C^\infty_c(D_{\psi_1,\psi_2})$, with $D_{\psi_1,\psi_2}:= {\rm
Dom}(\tilde{\psi}_1) \cap \tilde{\psi}_1^{-1}({\rm Dom}
(\tilde{\psi}_2)) \subset F^+$, maps $p \in D_{\psi_1,\psi_2}$ into
$f_1(p) \,f_2 (\tilde{\psi}_1(p)) \in {\bb R}$ (or ${\bb C})$. The
star on $U_\psi^*$ refers to the contravariant multiplication rule $
U^*_{\psi_1} U^*_{\psi_2} = U^*_{\psi_2 \psi_1}$. Associativity of
${\cal A}$ follows -- for appropriate support of the functions -- from
\[
\big(f_1 (f_2 \circ \psi_1)\big) (f_3 \circ (\psi_2\psi_1)) 
= f_1 \big((f_2 (f_3 \circ \psi_2)) \circ \psi_1 \big)~.
\]

We consider now the action of the vertical and horizontal vector
fields $Y^j_i$ and $X_i$ described in section \ref{general} on the
algebra ${\cal A}$. That action is simply defined as the action of the
vector fields on the functions,
\begin{equation}
Y^j_i ( f U_\psi^*) = Y^j_i (f) U_\psi^*~,\qquad 
X_i ( f U_\psi^*) = X_i (f) U_\psi^*~.
\end{equation}
The interesting effects we are looking for are obtained by application
of these vector fields to the product (\ref{prod}). For any vector
field $V$ on $F^+$ we compute 
\begin{eqnarray}
V(f_1 U^*_{\psi_1} \; f_2 U^*_{\psi_2}) \big|_p
&=& V(f_1 (f_2 \circ \tilde{\psi}_1)) U^*_{\psi_2 \psi_1} \big|_p
\nonumber \\
&=& \{V(f_1)  \big|_p\; (f_2 \circ \tilde{\psi}_1) \big|_p 
+ f_1 \big|_p \; V (f_2 \circ \tilde{\psi}_1) \big|_p\} 
U^*_{\psi_2 \psi_1}\nonumber  \\
&=& V(f_1) U^*_{\psi_1} \big|_p\; f_2 U^*_{\psi_2} 
+ f_1\big|_p \; ((\tilde{\psi}_{1*} V) f_2 )
\big|_{\tilde{\psi}_1(p)} U^*_{\psi_2 \psi_1} \nonumber \\
&=& V(f_1) U^*_{\psi_1} \big|_p\; f_2 U^*_{\psi_2} 
+ f_1\big|_p \; U^*_{\psi_1} U^*_{\psi_1^{-1}} 
((\tilde{\psi}_{1*} V) f_2) \big|_{\tilde{\psi}_1(p)} 
U^*_{\psi_2 \psi_1} \nonumber \\
&=& V(f_1) U^*_{\psi_1} \big|_p\; f_2 U^*_{\psi_2} 
+ f_1 U^*_{\psi_1} \; 
((\tilde{\psi}_{1*} V)  f_2 )\big|_{\tilde{\psi}_1^{-1}
\circ \tilde{\psi}_1(p)}  U^*_{\psi_2} \nonumber \\
&=& V(f_1 U^*_{\psi_1}) \big|_p\; f_2 U^*_{\psi_2} 
+ f_1 U^*_{\psi_1} \; \big(\tilde{\psi}_{1*}
(V\big|_{\tilde{\psi}_1^{-1}(p)}) \big) f_2 U^*_{\psi_2} \big|_p~.
\label{Vab}
\end{eqnarray}
In the third line we have used the definition of the push-forward. In
the fifth line we have commuted $U^*_{\psi_1^{-1}}$ with the function
$(\tilde{\psi}_{1*} V) f_2$, evaluated at $\tilde{\psi}_1(p)$.
According to (\ref{prod}), after taking $U^*_{\psi_1^{-1}}$ to the
right we must evaluate the function $(\tilde{\psi}_{1*} V) f_2$ at
$\tilde{\psi}_1^{-1}(\tilde{\psi}_1(p))=p$. This means that the
original field $V$ to push forward must be taken at
$\tilde{\psi}_1^{-1}(p)$. 

Taking for $V$ the vertical vector fields $Y^j_i$ and recalling their
invariance under diffeomorphisms (\ref{Ainv}), we obtain immediately 
\begin{equation}
Y^j_i(ab) = Y^j_i(a) \,b + a\, Y^j_i(b) ~,\qquad a,b \in {\cal A}~.
\label{Yab}
\end{equation}
The behavior of the horizontal vector fields $X_i$ is very different,
because they do not commute with the diffeomorphisms. Eq.\ 
(\ref{tX}) tells us that if $X_i$ is horizontal to $\omega$, then 
$X_i^{(\psi_1)}:=\tilde{\psi}_{1*}(X_i\big|_{\tilde{\psi}_1^{-1}(p)})$ is
horizontal to $(\tilde{\psi}^{-1}_1)^*\omega$. We denote the
connection coefficients of $(\tilde{\psi}^{-1}_1)^*\omega$ by
$\hat{\Gamma}^\nu_{\alpha\mu}$. We observe from (\ref{X}) and
(\ref{Y}) that
\begin{eqnarray}
(X_i^{(\psi_1)} -X_i )\big|_p = 
(\Gamma^\nu_{\alpha\mu}\big|_x -\hat{\Gamma}^\nu_{\alpha\mu}\big|_x)
y^\mu_i y^\alpha_j \partial^j_\nu &=&
(\Gamma^\nu_{\alpha\mu}\big|_x -\hat{\Gamma}^\nu_{\alpha\mu}\big|_x)
y^\mu_i y^\alpha_j (y^{-1})^k_\nu Y^j_k|_p \nonumber \\
&=:& 
\hat{\gamma}^k_{ji}\big|^{(\psi_1)}_p Y^j_k\big|_p~.
\end{eqnarray}
This gives from (\ref{Vab}) for the horizontal fields $X_i$
\begin{eqnarray}
X_i(f_1 U^*_{\psi_1} \; f_2 U^*_{\psi_2}) \big|_p
&=& X_i(f_1 U^*_{\psi_1}) \big|_p \; f_2 U^*_{\psi_2} \big|_p +
f_1 U^*_{\psi_1}\big|_p  \; X_i^{(\psi_1)} (f_2 U^*_{\psi_2})\big|_p  
\nonumber \\
&=& X_i(f_1 U^*_{\psi_1}) \big|_p \; f_2 U^*_{\psi_2} \big|_p +
f_1 U^*_{\psi_1} \big|_p \; X_i (f_2 U^*_{\psi_2}) \big|_p 
\nonumber \\
&& + f_1 U^*_{\psi_1} \big|_p \; \hat{\gamma}^k_{ij} 
\big|^{(\psi_1)}_p Y^j_k (f_2 U^*_{\psi_2}) \big|_p \nonumber 
\\
&=& X_i(f_1 U^*_{\psi_1}) \big|_p \; f_2 U^*_{\psi_2} \big|_p +
f_1 U^*_{\psi_1} \big|_p \; X_i (f_2 U^*_{\psi_2}) \big|_p 
\nonumber \\
&& + f_1 \big|_p \hat{\gamma}^k_{ij} \big|^{(\psi_1)}_{\tilde{\psi}_1(p)}
U^*_{\psi_1} \; Y^j_k (f_2 U^*_{\psi_2}) \big|_p ~.
\label{Xab}
\end{eqnarray}
Our goal is to express $\hat{\gamma}^k_{ij}
\big|^{(\psi_1)}_{\tilde{\psi}_1(p)}$ in terms of some function
evaluated at $p$. From (\ref{alpha}) and (\ref{om}) we conclude 
\begin{equation}
\omega^k_j \big|_p - ((\tilde{\psi}^{-1})^*\omega^k_j) \big|_p 
= \hat{\gamma}^k_{ji} \big|^{(\psi)}_p\; \alpha^i\big|_p~.
\label{hatgamma}
\end{equation}
We take this identity at $\tilde{\psi}(p)$ and apply $\tilde{\psi}^*$,
which gives
\begin{equation}
(\tilde{\psi}^* \omega^k_j) \big|_p - \omega^k_j \big|_p 
= \hat{\gamma}^k_{ji} \big|^{(\psi)}_{\tilde{\psi}(p)}\; 
(\tilde{\psi}^* \alpha^i) \big|_p
= \hat{\gamma}^k_{ji} \big|^{(\psi)}_{\tilde{\psi}(p)}\; 
\alpha^i \big|_p~,
\label{hgp}
\end{equation}
using the invariance of $\alpha^i$ under diffeomorphisms in the last
step. Replacing in (\ref{hgp}) $\psi$ by $\psi^{-1}$ and comparing with
(\ref{hatgamma}) we get
\begin{equation}
\hat{\gamma}^k_{ji} \big|^{(\psi)}_{\tilde{\psi}(p)} 
= - \hat{\gamma}^k_{ji} \big|^{(\psi^{-1})}_p 
=: \gamma^k_{ji} \big|^{(\psi)}_p = 
(\tilde{\Gamma}^\nu_{\alpha\mu}\big|_x -
\Gamma^\nu_{\alpha\mu}\big|_x) y^\alpha_j y^\mu_i (y^{-1})^k_\nu~,
\label{gamma}
\end{equation}
where $\tilde{\Gamma}^\nu_{\alpha\mu}$ and $\Gamma^\nu_{\alpha\mu}$
are the connection coefficients of the connections
$\tilde{\psi}^*\omega$ and $\omega$, respectively. Since
$\tilde{\Gamma}^\nu_{\alpha\mu}$ is defined by the diffeomorphism
$\psi$, we define an operator $\delta^k_{ji}$ on ${\cal A}$
by
\begin{equation}
\delta^k_{ji} (f U^*_\psi)\big|_{p} = \gamma^k_{ji}\big|^{(\psi)}_p\; 
f U^*_\psi\big|_{p} 
\label{deltaf}
\end{equation}
and get from (\ref{Xab}) and (\ref{gamma}) 
\begin{equation}
X_i(ab) = X_i(a)\,b + a\,X_i(b) + \delta^k_{ji}(a)\,Y^j_k(b)~,\qquad
a,b \in {\cal A}~.
\label{Xiab}
\end{equation}

Next, we compute 
\begin{equation}
\delta^k_{ji} (f_1 U^*_{\psi_1} \; f_2 U^*_{\psi_2}) \big|_p = 
\delta^k_{ji}(f_1 (f_2 \circ \tilde{\psi}_1) U^*_{\psi_2\psi_1}) \big|_p
= \gamma^k_{ji}\big|^{(\psi_2\psi_1)}_p \,
f_1\big|_p \,f_2\big|_{\tilde{\psi}_1(p)} U^*_{\psi_2\psi_1}~.
\label{deltaab}
\end{equation}
Starting with (\ref{hgp}) and (\ref{gamma}) we compute
\begin{eqnarray*}
\gamma^k_{ji}\big|^{(\psi_2\psi_1)}_p \;\alpha^i \big|_p 
&=& (\tilde{\psi}_2\tilde{\psi}_1)^*(\omega^k_j
\big|_{(\tilde{\psi}_2\tilde{\psi}_1)(p)}) - \omega^k_j\big|_p
\\
&=& \tilde{\psi}_1^* \Big( 
\tilde{\psi}_2^* (\omega^k_j 
\big|_{(\tilde{\psi}_2\tilde{\psi}_1)(p)}) 
-\omega^k_j \big|_{\tilde{\psi}_1(p)} \Big) 
+ \Big( \tilde{\psi}_1^*
(\omega^k_j\big|_{\tilde{\psi}_1(p)} ) - \omega^k_j \big|_p \Big)
\\
&=& \psi_1^* \Big(\gamma^k_{ji}\big|^{(\psi_2)}_{\tilde{\psi}_1(p)} 
\alpha^i \big|_{\tilde{\psi}_1(p)} \Big)
+ \gamma^k_{ji} \big|^{(\psi_1)}_p \alpha^i\big|_p
\\
&=& \Big( \gamma^k_{ji}\big|^{(\psi_1)}_p 
+ \gamma^k_{ji} \big|^{(\psi_2)}_{\tilde{\psi_1}(p)} \Big)
\alpha^i\big|_p~.
\end{eqnarray*}
We used again the invariance of $\alpha^i$ under diffeomorphisms in
the last line.  We insert this result into (\ref{deltaab}) and get
\begin{eqnarray*}
\delta^k_{ji} (f_1 U^*_{\psi_1} \; f_2 U^*_{\psi_2}) \big|_p 
&=& \gamma^k_{ji}\big|^{(\psi_1)}_p f_1\big|_p \,
f_2\big|_{\tilde{\psi}_1(p)} U^*_{\psi_2 \psi_1} 
+ f_1\big|_p \, \gamma^k_{ji}\big|^{(\psi_2)}_{\tilde{\psi}_1(p)}\,
f_2\big|_{\tilde{\psi}_1(p)} U^*_{\psi_2 \psi_1} 
\\
&=& \gamma^k_{ji}\big|^{(\psi_1)}_p f_1\big|_p U^*_{\psi_1}\; 
 f_2\big|_p U^*_{\psi_2}
+ f_1\big|_p U^*_{\psi_1} \;\gamma^k_{ji}\big|^{(\psi_2)}_p 
f_2\big|_p U^*_{\psi_2}~,
\end{eqnarray*}
which means 
\begin{equation}
\delta^k_{ji}(ab) = \delta^k_{ji}(a)\,b + a\,\delta^k_{ji}(b)~.
\label{delta}
\end{equation}

The equations (\ref{Yab}), (\ref{Xiab}) and (\ref{delta}) endow the
operators $X_i,Y_k^j$ and $\delta^k_{ji}$ with the structure of a
coalgebra, with the coproduct (\ref{gendelta}) given by
\begin{eqnarray}
\Delta(Y^j_k) &=& Y^k_j \otimes 1 + 1 \otimes Y^j_k~, \nonumber 
\\
\Delta(X_i) &=& X_i \otimes 1 + 1 \otimes X_i + \delta^k_{ji} \otimes 
Y^j_k~, \label{Delta}
\\
\Delta(\delta^k_{ji}) &=& \delta^k_{ji} \otimes 1 
+ 1 \otimes \delta^k_{ji}~, \nonumber 
\\
\Delta(1) &=& 1 \otimes 1~, \nonumber
\end{eqnarray}
with $1$ being the identity on ${\cal A}$. It is easy to check that
$\Delta$ is coassociative on the linear space 
${\bb R}(1,X_i,Y_k^j,\delta^k_{ji})$,
\begin{equation}
(\Delta \otimes {\rm id}) \circ \Delta = ({\rm id} \otimes \Delta) 
\circ \Delta~. 
\label{coass}
\end{equation}

\section{From Lie algebra to Hopf algebra}
\label{Lie}

Vector fields form a Lie algebra, so it is natural to investigate
whether $X_i,Y_k^j,\delta^k_{ji}$ generate a Lie algebra. We compute
the mutual commutators, starting with $Y^i_j$:
\begin{eqnarray}
[Y^i_j,Y^k_l] (fU^*_\psi) &=& (y^\mu_j \partial^i_\mu 
y^\nu_l \partial^k_\nu -  y^\nu_l \partial^k_\nu y^\mu_j
\partial^i_\mu ) fU^*_\psi 
\nonumber 
\\
&=& (\delta^i_l Y^k_j - \delta^k_j Y^i_l) (fU^*_\psi)~,
\\{}
[Y^k_j,X_i] (fU^*_\psi) &=& \big( y^\mu_j \partial^k_\mu 
(y^\nu_i \partial_\nu - \Gamma^\beta_{\alpha\nu} y^\nu_i y^\alpha_l 
\partial^l_\beta) 
-(y^\nu_i \partial_\nu - \Gamma^\beta_{\alpha\nu} y^\nu_i y^\alpha_l 
\partial^l_\beta) y^\mu_j \partial^k_\mu\big) fU^*_\psi  
\nonumber
\\
&=& \delta^k_i X_j (fU^*_\psi)~,
\\{}
[Y^i_j,\delta^k_{lm}] (fU^*_\psi) &=& \big( y^\mu_j \partial^i_\mu 
(\tilde{\Gamma}^\nu_{\beta\alpha}{-} \Gamma^\nu_{\beta\alpha}) 
y^\beta_l y^\alpha_m (y^{-1})^k_\nu - 
(\tilde{\Gamma}^\nu_{\beta\alpha}{-}\Gamma^\nu_{\beta\alpha}) 
y^\beta_l y^\alpha_m (y^{-1})^k_\nu y^\mu_j \partial^i_\mu \big)
fU^*_\psi \nonumber
\\
&=& (\delta^i_l \delta^k_{jm} + \delta^i_m \delta^k_{lj} - 
\delta^k_j \delta^i_{lm}) ( fU^*_\psi)~.
\label{Ydel}
\end{eqnarray}

So far we have considered the most general connection on $M$, even
with torsion. But now, the commutator of horizontal vector fields
\begin{eqnarray}
[X_i,X_j] &=& R^k_{lij} Y^l_k + \Theta^k_{ij} X_k~, \\
R^k_{lij} &=& (y^{-1})^k_\sigma y^\rho_l y^\mu_i y^\nu_j \big( 
\partial_\nu\Gamma^\sigma_{\rho\mu}-
\partial_\mu\Gamma^\sigma_{\rho\nu} + \Gamma^\beta_{\rho\mu}
\Gamma^\sigma_{\beta\nu} - \Gamma^\beta_{\rho\nu}
\Gamma^\sigma_{\beta\mu} \big)~, \nonumber \\
\Theta^k_{ij} &=& (y^{-1})^k_\rho y^\mu_i y^\nu_j \big( 
\Gamma^\rho_{\mu\nu} - \Gamma^\rho_{\nu\mu}\big)~, \nonumber
\end{eqnarray}
leads to curvature $R$ and torsion $\Theta$, i.e.\ not to structure
`constants'. Torsion can be avoided by the choice of the connection,
but we would be forced to include $R^k_{lij} Y^l_k$ and its repeated
commutators with $X_m$ in the list of generators of the Lie algebra we
are looking for. To avoid these terms we follow \cite{cm2} and
restrict ourselves to a {\it flat} manifold. Locally this is always
possible, and globally it is achieved via the {\it Morita
equivalence}. For a locally finite cover of the manifold $M$ by charts
$U_\alpha$, let $N=\coprod U_\alpha$ be the disjoint union of the
charts. Moreover, let $\Gamma'$ be the pseudogroup of local
diffeomorphisms of $N$. Without giving the proof we recall from
\cite{cm2} that the two algebras ${\cal A} = C^\infty_c(F^+(M)) \semi
\Gamma$ and ${\cal A}' = C^\infty_c(F^+(N)) \semi \Gamma'$ are Morita
equivalent. There is a canonical connection on $N$, the flat
connection given by $\Gamma^\alpha_{\beta\gamma}=0$. This means that
given $M$ we pass to $N$ and the corresponding crossed product ${\cal
A}'$ and derive there the coproduct and Lie algebra structure of
vector fields on $F^+(N)$ for the flat connection.

Thus, the horizontal vector fields take the simple form $X_i = y^\mu_i
\partial_\mu$, and they now commute with each other:
\begin{equation}
[X_i,X_j] (f U^*_\psi) = 0 ~.
\end{equation}
Due to (\ref{Gamma}), (\ref{gamma}) and (\ref{deltaf}), the action of
$\delta^k_{ji}$ on ${\cal A}$ simplifies in the case of a flat
manifold to 
\begin{equation}
\delta^k_{ji} (f U^*_\psi) = ((\partial\psi(x))^{-1})^\nu_\beta\,
\partial_\mu\partial_\alpha \psi(x)^\beta \, y^\mu_j y^\alpha_i
(y^{-1})^k_\nu\, f U^*_\psi~.
\end{equation}
The (repeated) commutator with $X_l$ leads to new operators on ${\cal
A}$,
\begin{eqnarray}
\delta^k_{ji,l_1\dots l_n} (f U^*_\psi) &:=& 
[X_{l_n},\dots ,[X_{l_1},\delta^k_{ji}]\dots] (f U^*_\psi) 
\label{flat}
\\ \nonumber 
&=& \partial_{\lambda_n} \dots \partial_{\lambda_1} 
\Big(((\partial\psi(x))^{-1})^\nu_\beta\,
\partial_\mu\partial_\alpha \psi(x)^\beta \Big) y^\mu_j y^\alpha_i
(y^{-1})^k_\nu\, y^{\lambda_1}_{l_1} \cdots y^{\lambda_n}_{l_n} \; 
f U^*_\psi~.
\end{eqnarray}
It is clear that all these operators $\delta$ commute with each other,
\begin{equation}
[\delta^k_{ji,l_1\dots l_n},\delta^c_{ba,d_1\dots d_n}](f U^*_\psi)
=0~.
\end{equation}

We see that the linear space generated by $X_i,Y^k_j,
\delta^k_{ji,l_1\dots l_n}$ forms a Lie algebra, and we let ${\cal H}$
be the corresponding enveloping algebra. This is the algebra of
polynomials in the generators of the Lie algebra, with the commutation
relations inherited from the Lie algebra. Thus a
(Poincar\'e-Birkhoff-Witt) basis in ${\cal H}$ is given by
\[
X_{i_1} \cdots X_{i_\alpha} Y^{k_1}_{j_1} \cdots 
Y^{k_\beta}_{j_\beta} \delta^{a_1}_{b_1 c_1} \cdots  
\delta^{a_\gamma}_{b_\gamma c_\gamma} \delta^{d_1}_{e_1 f_1,h_1} \cdots  
\delta^{d_\delta}_{e_\delta f_\delta,h_\delta} \cdots  ~,
\]
with $i_1\leq i_2\leq \dots \leq i_n$ and so on for the other
indices. We extend the coproduct (\ref{Delta}) recursively to ${\cal
H}$ by the definition
\begin{equation}
\Delta (h^1 h^2) = \Delta (h^1) \, \Delta(h^2) := \sum h_1^1 h^1_2 \otimes 
h_1^2 h^2_2 ~,\qquad \Delta(h_i) = \sum h^1_i \otimes h^2_i~,
\label{copro}
\end{equation}
for $h_1,h_2 \in {\cal H}$. The coproduct is automatically
coassociative (\ref{coass}) and by construction (\ref{copro})
compatible with the multiplication in ${\cal H}$. 

For notational convenience we abbreviate $\delta^A=\delta^k_{ji}$ with
$A=1,\dots, n^2(n+1)/2$, due to symmetry in $i,j$. Moreover, we
introduce a string $a=a_1a_2\dots a_k$ for the repeated commutators
with $X_{a_1},\dots,X_{a_k}$ and denote its length by $|a|=k$.  Next,
let ${\cal H}_n$ be the commutative algebra of polynomials in the
variables $1$ and $\delta^A_a$, with $0\leq |a| \leq n$. Let ${\cal
H}^0_n$ be the ideal of polynomials vanishing at $0$. We obtain a more
explicit formula of the coproduct in 
\begin{Lemma}
$\Delta \delta^A_a = \delta^A_a \otimes 1 + 1 \otimes \delta^A_a +
R^A_a~, \qquad R^A_a \in {\cal H}^0_{n-1} \otimes {\cal H}^0_{n-1}
\quad \mbox{for} ~~ |a|=n~.$
\label{dex}
\end{Lemma}
{\it Proof.} The Lemma holds for $n=0$ with $R^A=0$. Assuming it holds
for $|a|=n$ we compute for $b:=ai$ (appending the index $i$ to the
string $a$), $|b|=n+1$, 
\begin{eqnarray}
\Delta(\delta^A_b) &=& \Delta([X_i,\delta^A_a]) =
[ \Delta(X_i),\Delta(\delta^A_a)] \nonumber 
\\
&=& [X_i \otimes 1 + 1 \otimes X_i + \delta^k_{ji} \otimes Y^j_k , 
\delta^A_a \otimes 1 + 1 \otimes \delta^A_a + R^A_a] \nonumber
\\
&=& \delta^A_b \otimes 1 + 1 \otimes \delta^A_b + R^A_b~,
\quad\mbox{with} \nonumber
\\[1ex]
R^A_{ai} &:=& [X_i \otimes 1 + 1 \otimes X_i + \delta^k_{ji} \otimes
Y^j_k, R^A_a] + \delta^k_{ji} \otimes [Y^j_k,\delta^A_a]~.
\label{R}
\end{eqnarray}
For $n=1$ we get $R^A_i = \delta^k_{ji}  \otimes [Y^j_k,
\delta^A] \in {\cal H}^0_0 \otimes {\cal H}^0_0$. The Lemma follows
from the fact that the commutator with $Y^j_k$ preserves ${\cal
H}^0_m$ whereas the commutator with $X_i$ sends elements of ${\cal
H}^0_m$ to elements of ${\cal H}^0_{m+1}$. \qed
\vskip 2ex

\noindent
For example, we obtain from (\ref{Ydel}) immediately
\begin{equation}
\Delta(\delta^k_{ji,l})=\delta^k_{ji,l} \otimes 1 + 1 \otimes 
\delta^k_{ji,l} + \delta^a_{jl} \otimes \delta^k_{ai}
+ \delta^a_{il}  \otimes \delta^k_{ja} 
- \delta^k_{al} \otimes  \delta^a_{ji}~.
\label{d2}
\end{equation}

The counit $\epsilon$ on ${\cal H}$ is defined by 
\begin{equation}
\varepsilon(1)=1~,\qquad \varepsilon(h)=0\quad 
\forall h\neq 1 ~.
\end{equation}
The counit axiom 
\[
(\varepsilon \otimes {\rm id}) \circ \Delta(h) = 
({\rm id} \otimes \varepsilon) \circ \Delta(h) = h\qquad \forall\,
h\in {\cal H} 
\]
is clear for $h=X_i,Y^k_j,\delta^A$. For $\delta^A_a$ it follows from
Lemma \ref{dex}, using $\varepsilon (h^0)=0$ for $h^0 \in {\cal H}_n^0$.

Therefore, ${\cal H}$ is a bialgebra
(algebra+coalgebra+compatibility), and our next task is to show the
existence of an antipode $S$ on ${\cal H}$, making ${\cal H}$ to a
Hopf algebra. The antipode has to satisfy the axioms
\begin{eqnarray}
S(h_1 h_2) = S(h_2) S(h_1)~,
\nonumber \\
m \circ (S \otimes {\rm id}) \circ \Delta(h) = \varepsilon(h)~,
\label{Sax}
\\ \nonumber
m \circ ({\rm id} \otimes S) \circ \Delta(h) = \varepsilon(h)~,
\end{eqnarray}
for $h,h_1,h_2 \in {\cal H}$, and where $m$ denotes the
multiplication. Applying (\ref{Sax}) to $1,Y^j_k,\delta^k_{ji},X_i
\in {\cal H}$, in that order, we get
\begin{eqnarray}
S(1) &=& 1 ~, \nonumber \\
S(Y^j_k) &=& - Y^j_k~, \nonumber \\
S(\delta^k_{ji}) &=& - \delta^k_{ji}~, \\
S(X_i) &=& - X_i + \delta^k_{ji} Y^j_k~, \nonumber
\end{eqnarray}
The antipode on $\delta^A_a$ is obtained from (\ref{Sax}) by recursion
in $|a|$, with the task to prove that the tree possible definitions
coincide. First, employing the Sweedler notation 
$\Delta(R_a)=R_{a(1)} \otimes R_{a(2)}$ (and omitting the summation sign),
we have with (\ref{R}) 
\begin{eqnarray}
S(\delta^A_{ai}) &=& -\delta^A_{ai} - m \circ (S \otimes {\rm
id}) \circ \Delta(R^A_{ai}) \nonumber \\*
&=&  -\delta^A_{ai} - S ([X_i,R^A_{a(1)}]) \,R^A_{a(2)} 
- S (R^A_{a(1)}) \,[X_i, R^A_{a(2)}] 
- S (\delta^k_{ji} R^A_{a(1)}) \,[Y^j_k, R^A_{a(2)}] 
\nonumber \\*
&& {}\qquad - S(\delta^k_{ji}) [Y^j_k,\delta^A_a] 
\nonumber \\
&=&  -\delta^A_{ai}  - [X_i-\delta^k_{ji} Y^j_k,S(R^A_{a(1)})] \,
R^A_{a(2)} - S (R^A_{a(1)}) \,[X_i, R^A_{a(2)}] 
\nonumber \\
&& {}\qquad
+ \delta^k_{ji} S(R^A_{a(1)}) \,[Y^j_k, R^A_{a(2)}] 
+ \delta^k_{ji} [Y^j_k,\delta^A_a] 
\nonumber \\
&=&  -\delta^A_{ai}  + [-X_i + \delta^k_{ji} Y^j_k, 
S (R^A_{a(1)}) R^A_{a(2)}] + \delta^k_{ji} [Y^j_k,\delta^A_a] 
\nonumber \\
&=& [S(\delta^A_a),-X_i+\delta^k_{ji} Y^j_k]=[S(\delta^A_a),S(X_i)]~.
\end{eqnarray}
In the same way one checks $-\delta^A_{ai} - m \circ ({\rm id} \otimes
S) \circ \Delta (R^A_{ai}) = [S(\delta^A_a),S(X_i)]$. For example, one
easily obtains
\begin{equation}
S(\delta^k_{ji,l}) = -  \delta^k_{ji,l} 
+ \delta^a_{jl} \delta^k_{ai} + \delta^a_{il} \delta^k_{ja} 
- \delta^k_{al} \delta^a_{ji}~.
\label{Sd2}
\end{equation}

This finishes our review of the construction of the Connes--Moscovici
Hopf algebra \cite{cm2}. In their work, the cyclic cohomology of
this Hopf algebra serves as an organizing principle for the
computation of the cocycles in the local index formula \cite{cm1}. We
hope to be more specific on that point in the future.

\section{Explicit solution: rooted trees}

Following an idea by Connes and Kreimer \cite{ck} we will now describe
the commutative Hopf algebra ${\cal H}_n$ of polynomials in
$\delta^A_a$, $|a| \leq n$, by graphical tools, generalized from the
one-dimensional case in \cite{ck} to arbitrary dimension of the
manifold $M$. In this way we obtain a Hopf algebra of rooted trees,
which is intimately related to a Hopf algebra structure in
perturbative quantum field theories as discovered by Kreimer
\cite{k1}. The antipode of Kreimer's Hopf algebra achieves the
renormalization of divergent Feynman graphs, see \cite{k1,k2}. 

We label the generator $\delta^k_{ji}$ by an indexed dot,
\begin{equation}
\delta^k_{ji} = ~ \bullet~^k_{ji}~.  
\end{equation}
The goal is to derive the symbol for $\delta^k_{ji,l}$. This goes via
the coproduct (\ref{d2}), which tells us after comparison with (\ref{R})
\begin{equation}
\bullet~^a_{bl} \otimes [Y^b_a, ~\bullet~^k_{ji}] = 
{} ~\bullet~^a_{jl} \otimes \bullet~^k_{ai} ~
+ ~\bullet~^a_{il} \otimes \bullet~^k_{ja}~ 
- ~\bullet~^k_{bl} \otimes \bullet~^b_{ji} ~.
\label{b2}
\end{equation}
The commutator with $Y$ picks up one index of $~\bullet~^k_{ji}$ and
moves it to the first upper or lower place in $~\bullet~^a_{bl}~$,
overwriting the index there. The vacant position in $~\bullet~^k_{ji}$
is filled with the remaining summation index of
$~\bullet~^a_{bl}~$. If the index picked up was a lower (upper) one,
we count the resulting tensor product positive (negative). This leads
us to think of the rhs of (\ref{b2}) as being produced by a cut of a
symbol
\[
\delta^k_{ji,l} = \parbox{8mm}{\begin{picture}(6,11)
\put(1,8){$\bullet~^k_{ji}$}
\put(2,9){\line(0,-1){7}}
\put(1,1){$\bullet~_l$}
\end{picture}} \quad \longrightarrow \quad 
\parbox{8mm}{\begin{picture}(6,11)
\put(1,8){$\bullet~^k_{ji}$}
\put(2,9){\line(0,-1){7}}
\put(0,4){---}
\put(1,1){$\bullet~_l$}
\end{picture}} 
= \parbox{8mm}{\begin{picture}(6,11)
\put(1,8){$\bullet~^k_{ai}$}
\put(1,1){$\bullet~^a_{jl}$}
\end{picture}} 
+ \parbox{8mm}{\begin{picture}(6,11)
\put(1,8){$\bullet~^k_{ja}$}
\put(1,1){$\bullet~^a_{il}$}
\end{picture}} 
- \parbox{8mm}{\begin{picture}(6,11)
\put(1,8){$\bullet~^a_{ij}$}
\put(1,1){$\bullet~^k_{al}$}
\end{picture}} ~.
\]
We call the uppermost index which is different from the lower index
the root. The graph above the cut connected with the root is called
the trunk and goes to the rhs of the tensor product. A graph below the
cut is called a cut branch and goes to the lhs of the tensor
product. We define the action of a cut as the movement of one index of
the vertex above the cut to the first position of the new root of the
cut branch. The remaining position to complete the root of the cut
branch is filled with a summation index and the same summation index
is put into the vacant position of the trunk. In the case of cutting
immediately below the root, we have to sum over the three
possibilities of picking up indices of the root, adding a minus sign
if we pick up the unique upper index. We thus get the following
graphical interpretation of (\ref{d2}):
\begin{equation}
\Delta \left( \parbox{8mm}{\begin{picture}(6,11)
\put(1,8){$\bullet~^k_{ji}$}
\put(2,9){\line(0,-1){7}}
\put(1,1){$\bullet~_l$}
\end{picture}}  \right)
 = \left[ \parbox{8mm}{\begin{picture}(6,11)
\put(1,8){$\bullet~^k_{ji}$}
\put(2,9){\line(0,-1){7}}
\put(1,1){$\bullet~_l$}
\end{picture}}  \right]^c
+ 
\left[ \parbox{8mm}{\begin{picture}(6,11)
\put(1,8){$\bullet~^k_{ji}$}
\put(2,9){\line(0,-1){7}}
\put(1,1){$\bullet~_l$}
\end{picture}} \right]_c + ~
\parbox{8mm}{\begin{picture}(6,11)
\put(1,8){$\bullet~^k_{ji}$}
\put(2,9){\line(0,-1){7}}
\put(0,4){---}
\put(1,1){$\bullet~_l$}
\end{picture}} ~.
\label{db2}
\end{equation}
On the rhs, $[\delta]^c$ stands for $\delta \otimes 1$ (cutting above
the entire tree) and $[\delta]_c$ for $1 \otimes \delta$ (cutting
below the entire tree). 

The next step is to compute $\Delta(\delta^k_{ji,lm})$ by commuting
$\Delta(X_m)$ with (\ref{db2}). The term $[\delta^k_{ji,l}]^c$ has a
non-vanishing commutator only with $X_m \otimes 1$. It yields
$\delta^k_{ji,lm} \otimes 1$, and this trivial behavior continues to
higher degrees. Next, $X_m \otimes 1$ commutes with $[\delta]_c$, whereas
\begin{equation}
\left[ X_m \otimes 1 , \parbox{8mm}{\begin{picture}(6,11)
\put(1,8){$\bullet~^k_{ji}$}
\put(2,9){\line(0,-1){7}}
\put(0,4){---}
\put(1,1){$\bullet~_l$}
\end{picture}} \right] 
= \parbox{8mm}{\begin{picture}(6,18)
\put(1,15){$\bullet~^k_{ji}$}
\put(2,16){\line(0,-1){14}}
\put(0,11){---}
\put(1,8){$\bullet~_l$}
\put(1,1){$\bullet~_m$}
\end{picture}} 
= \delta^a_{jl,m} \otimes \delta^k_{ai} + 
\delta^a_{il,m} \otimes \delta^k_{ja} - \delta^k_{al,m} \otimes
\delta^a_{ji} ~.
\label{lm}
\end{equation}
Our previous definition of a cut extends without modification to that
case. The term $1 \otimes X_m$ commuted with $[\delta^k_{ji,l}]_c$
gives $[\delta^k_{ji,lm}]_c$, whereas 
\begin{equation}
\left[ 1 \otimes X_m , \parbox{8mm}{\begin{picture}(6,11)
\put(1,8){$\bullet~^k_{ji}$}
\put(2,9){\line(0,-1){7}}
\put(0,4){---}
\put(1,1){$\bullet~_l$}
\end{picture}} \right] 
= \parbox{14mm}{\begin{picture}(14,11)
\put(4.5,8){$\bullet~~^k_{ji}$}
\put(5.5,9){\line(-1,-2){3.5}}
\put(5.5,9){\line(1,-2){3.5}}
\put(1.5,4){---}
\put(1,1){$\bullet~_l$}
\put(8,1){$\bullet~_m$}
\end{picture}} = \delta^a_{jl} \otimes \delta^k_{ai,m} + 
\delta^a_{il} \otimes \delta^k_{ja,m} - \delta^k_{al} \otimes
\delta^a_{ji,m} ~.
\end{equation}
The cut on the tree in the middle only sees the indices $k,j,i$ -- but
not $m$ -- by the definition of a cut as affecting only the indices of
the unique vertex above the cut. With this rule we get easily the
corresponding expression in terms of $\delta$'s on the rhs. The
commutator of $\delta^c_{bm} \otimes Y^b_c$ with $[\delta^k_{ji,l}]_c$
moves the indices $k,j,i,l$ to their correct position in
$\delta^c_{bm}$, and this is precisely obtained as the sum of two
different cuts: 
\begin{eqnarray}
\left[ \delta^c_{bm} \otimes Y^b_c, \parbox{8mm}{\begin{picture}(6,11)
\put(1,8){$\bullet~^k_{ji}$}
\put(2,9){\line(0,-1){7}}
\put(1,1){$\bullet~_l$}
\end{picture}} \right] 
&=& \parbox{8mm}{\begin{picture}(6,18)
\put(1,15){$\bullet~^k_{ji}$}
\put(2,16){\line(0,-1){14}}
\put(0,4){---}
\put(1,8){$\bullet~_l$}
\put(1,1){$\bullet~_m$}
\end{picture}} 
+ 
\parbox{14mm}{\begin{picture}(14,11)
\put(4.5,8){$\bullet~^k_{ji}$}
\put(5.5,9){\line(-1,-2){3.5}}
\put(5.5,9){\line(1,-2){3.5}}
\put(5.5,4){---}
\put(1,1){$\bullet~_l$}
\put(8,1){$\bullet~_m$}
\end{picture}} ~,  \label{lm2}
\\[-2ex] \nonumber
\parbox{14mm}{\begin{picture}(14,11)
\put(4.5,8){$\bullet~^k_{ji}$}
\put(5.5,9){\line(-1,-2){3.5}}
\put(5.5,9){\line(1,-2){3.5}}
\put(5.5,4){---}
\put(1,1){$\bullet~_l$}
\put(8,1){$\bullet~_m$}
\end{picture}} &=& \delta^a_{jm} \otimes \delta^k_{ai,l} + 
\delta^a_{im} \otimes \delta^k_{ja,l} 
- \delta^k_{am} \otimes \delta^a_{ji,l}~,\quad 
\parbox{8mm}{\begin{picture}(6,18)
\put(1,15){$\bullet~^k_{ji}$}
\put(2,16){\line(0,-1){14}}
\put(0,4){---}
\put(1,8){$\bullet~_l$}
\put(1,1){$\bullet~_m$}
\end{picture}} =  \delta^a_{lm} \otimes \delta^k_{ji,a} ~.
\end{eqnarray}
There remains one final commutator to compute, that of 
$\delta^c_{bm} \otimes Y^b_c$ with the graph in
(\ref{db2}) already cut. For each of the tree terms corresponding
to the previous cut, we have to move each of the tree indices of its
root down to $\delta^c_{bm}$. This gives the following symbolic
expression of these nine tensor products: 
\begin{eqnarray}
\left[ \delta^c_{bm} \otimes Y^b_c, 
\parbox{8mm}{\begin{picture}(6,11)
\put(1,8){$\bullet~^k_{ji}$}
\put(2,9){\line(0,-1){7}}
\put(0,4){---}
\put(1,1){$\bullet~_l$}
\end{picture}} \right] 
= \parbox{14mm}{\begin{picture}(14,11)
\put(4.5,8){$\bullet~^k_{ji}$}
\put(5.5,9){\line(-1,-2){3.5}}
\put(5.5,9){\line(1,-2){3.5}}
\put(1.5,4){---}
\put(6,3){---}
\put(1,1){$\bullet~_l$}
\put(8,1){$\bullet~_m$}
\end{picture}}
&=& \delta^a_{jl} \delta^b_{am} \otimes \delta^k_{bi} + 
\delta^a_{jl} \delta^b_{im} \otimes \delta^k_{ab} - 
\delta^a_{jl} \delta^k_{bm} \otimes \delta^b_{ai} \nonumber
\\[-1ex]
&+& \delta^a_{il} \delta^b_{jm} \otimes \delta^k_{ba} 
+ \delta^a_{il} \delta^b_{am} \otimes \delta^k_{jb} 
- \delta^a_{il} \delta^k_{bm} \otimes \delta^b_{ja} \nonumber
\\[0.5ex]
&-& \delta^k_{al} \delta^b_{jm} \otimes \delta^a_{bi} 
- \delta^k_{al} \delta^b_{im} \otimes \delta^a_{jb} 
+ \delta^k_{al} \delta^a_{bm} \otimes \delta^b_{ji} ~.\quad{}
\label{cut2}
\end{eqnarray}
Note that the order of the cuts in this graph is important, we first
have to cut the vertex $l$ away and then the vertex $m$. 

Our construction leads us to define
\begin{equation}
\delta^k_{ji,lm} = \parbox{8mm}{\begin{picture}(6,18)
\put(1,15){$\bullet~^k_{ji}$}
\put(2,16){\line(0,-1){14}}
\put(1,8){$\bullet~_l$}
\put(1,1){$\bullet~_m$}
\end{picture}} + 
\parbox{14mm}{\begin{picture}(14,11)
\put(4.5,8){$\bullet~^k_{ji}$}
\put(5.5,9){\line(-1,-2){3.5}}
\put(5.5,9){\line(1,-2){3.5}}
\put(1,1){$\bullet~_l$}
\put(8,1){$\bullet~_m$}
\end{picture}} ~.
\end{equation}

\begin{Definition}
Let $\delta^A_a = \sum_{k=1}^{|a|!} t^{|a|}_k$ be recursively
represented by a sum of $|a|!$ connected rooted trees, each of
them having $|a|{+}1$ vertices. We define
\begin{equation}
\delta^A_{ai} \equiv [X_i,\delta^A_a] = \sum_{k=1}^{|a|!}
\sum_{j=1}^{|a|+1} t^{|a|}_{k_j} =: \sum_{\ell=1}^{|ai|!}
t^{|ai|}_{\ell}~,
\label{tkj}
\end{equation}
where the rooted tree $t^{|a|}_{k_j}$ is obtained by attaching the new 
vertex $i$ to the right of the $j^{\rm th}$ vertex of $t^{|a|}_k$. 
\end{Definition}

\begin{Proposition}
The coproduct of $\delta^A_a = \sum_{k=1}^{|a|!} t^{|a|}_k$ is given by
\begin{equation}
\Delta (\delta^A_a)= \delta^A_a \otimes 1+1 \otimes \delta^A_a + 
 \sum_{k=1}^{|a|!} \sum_{{\cal C}} P^{\cal C}(t^{|a|}_k) \otimes 
R^{\cal C}(t^{|a|}_k)~,
\label{PR}
\end{equation}
where for each $t^{|a|}_k$ the sum is over all admissible cuts ${\cal
C}$ of $t^{|a|}_k$ (i.e.\ those non-empty multiple cuts for which on
each path from the bottom to the root there is at most one individual
cut). In eq.\ (\ref{PR}), $R^{\cal C}(t^{|a|}_k)$ is the trunk and
$P^{\cal C}(t^{|a|}_k)$ the product of cut branches obtained by
cutting $t^{|a|}_k$ via the multiple cut ${\cal C}$. If immediately
below a vertex there are several cuts on outgoing edges, the order of
the cuts is from left to right.
\end{Proposition}
{\it Proof.}  Commuting $\Delta(X_i)$ with $\Delta(\delta^A_a)$ to get
$\Delta(\delta^A_{ai})$, the term $\delta^A_a \otimes 1$ develops into
$\delta^A_{ai} \otimes 1$. Next, $X_i \otimes 1$ attaches successively
a vertex $i$ to each vertex of the cut branches $P^{\cal
C}(t^{|a|}_k)$, and $1 \otimes X_i$ does the same for the trunk
$R^{\cal C}(t^{|a|}_k)$ of each tree $t^{|a|}_k$ constituting
$\delta^A_a$. Finally $\delta \otimes Y$ attaches a cut-away vertex
everywhere on the trunk, not on the cut branch. This excludes multiple
cuts on paths from bottom to top. The result clearly reproduces our
prescription of the coproduct of $\delta^A_{ai}$, see (\ref{tkj}). \qed
\vskip 2ex

We make one important observation. Although the operators $\delta$ are
invariant under permutation of the indices after the comma, for
instance $\delta^k_{ji,lm}=\delta^k_{ji,ml}$, see (\ref{flat}), this
symmetry is lost on the level of individual trees, see for instance
(\ref{lm}). However, these terms combined with the `diagonal' terms of 
(\ref{lm2}) are symmetric in $l$ and $m$. 

We recall that in Kreimer's Hopf algebra of renormalization
\cite{k1,k2} a rooted tree represents the divergence structure of a
Feynman graph.  A divergent sector in such a graph is represented
by a vertex. The root represents the overall (superficial)
divergence. The construction rule for the tree is -- in absence of
overlapping subdivergences -- to put subdivergences $\gamma_i$ of a
divergence $\gamma$ into down-going branches of $\gamma$. Disjoint
divergences are only indirectly connected via the divergence which
contains them as subdivergences.  Overlapping divergences have to be
resolved in terms of disjoint and nested ones and give a sum of trees,
see \cite{kw,k4}.

The $n$-dimensional case treated here is closer to quantum field
theory than dimension $1$ because we obtain \emph{decorated} trees --
the decoration here being given by spacetime indices (three for the
root) whereas in QFT it is a label for divergent Feynman graphs
without subdivergences. In this sense, a (not super-) renormalizable
QFT has something to do with diffeomorphisms on an infinite
dimensional manifold. Our observation leads us to speculate that {\it
the sum of Feynman graphs according to the collection of rooted trees
to $\delta$'s has more symmetry than the individual Feynman graphs.}
This should be checked in QFT calculations. Another interpretation
would be the observation
\begin{equation}
\parbox{8mm}{\begin{picture}(6,18)
\put(1,15){$\bullet~^k_{ji}$}
\put(2,16){\line(0,-1){14}}
\put(1,8){$\bullet~_l$}
\put(1,1){$\bullet~_m$}
\end{picture}} + 
\parbox{14mm}{\begin{picture}(14,11)
\put(4.5,8){$\bullet~^k_{ji}$}
\put(5.5,9){\line(-1,-2){3.5}}
\put(5.5,9){\line(1,-2){3.5}}
\put(1,1){$\bullet~_l$}
\put(8,1){$\bullet~_m$}
\end{picture}} - 
 \parbox{8mm}{\begin{picture}(6,18)
\put(1,15){$\bullet~^k_{ji}$}
\put(2,16){\line(0,-1){14}}
\put(1,8){$\bullet~_m$}
\put(1,1){$\bullet~_l$}
\end{picture}} -
\parbox{14mm}{\begin{picture}(14,11)
\put(4.5,8){$\bullet~^k_{ji}$}
\put(5.5,9){\line(-1,-2){3.5}}
\put(5.5,9){\line(1,-2){3.5}}
\put(1,1){$\bullet~_m$}
\put(8,1){$\bullet~_l$}
\end{picture}} =0 ~,
\label{rel}
\end{equation}
which could possibly be regarded as a relation between Feynman
graphs similar to those derived in \cite{k3}\footnote{Dirk Kreimer
confirmed to me that (\ref{rel}) is satisfied in QFT for the leading
divergences, as it can be derived from sec.\ V.C in \cite{kd}. For
non-leading singularities there will be (probably systematic)
modifications.}. 
\begin{Proposition}
The antipode $S$ of $\delta^A_a=\sum_{k=1}^{|a|!}
t^{|a|}_k$ is given by 
\begin{equation}
S(\delta^A_a) = - \delta^A_a - \sum_{k=1}^{|a|!} \sum_{{\cal C}_a} 
(-1)^{|{\cal C}_a|}\; P^{{\cal C}_a}(t^{|a|}_k)\,R^{{\cal C}_a}(t^{|a|}_k)~,
\label{Sprp}
\end{equation}
where the sum is over the set of all non-empty multiple cuts ${\cal
C}_a$ of $t^{|a|}_k$ (multiple cuts on paths from bottom to the root
are allowed) consisting of $|{\cal C}_a|$ individual cuts. The order
of cuts is from top to bottom and from left to right.
\end{Proposition}
{\it Proof.} We apply the antipode axiom $m \circ(S \otimes {\rm id})
\circ \Delta = 0$, see (\ref{Sax}), to (\ref{PR}), giving with
$S(1)=1$ the recursion
\[
S(\delta^A_a) = - \delta^A_a - \sum_{k=1}^{|a|!} \sum_{\cal C}
\Big( \prod_{j=1}^{|{\cal C}|} S( t^{|a|,{\cal C}}_{k,j}) \Big)\,
R^{\cal C}(t^{|a|}_k)~, \qquad 
P^{\cal C}(t^{|a|}_k)= \prod_{j=1}^{|{\cal C}|} t^{|a|,{\cal C}}_{k,j}~,
\]
where $|{\cal C}|$ is the number of individual cuts in ${\cal
C}$. For each $\{{\cal C},j\}$ we have 
\begin{equation}
S( t^{|a|,{\cal C}}_{k,j})= - t^{|a|,{\cal C}}_{k,j} - \sum_{{\cal
C}_j} S(P^{{\cal C}_j}(t^{|a|,{\cal C}}_{k,j})) 
R^{{\cal C}_j}(t^{|a|,{\cal C}}_{k,j})~,
\label{Cj}
\end{equation}
where the sum is over the set of admissible cuts ${\cal C}_j$ of
$t^{|a|,{\cal C}}_{k,j}$. In the first level, the product
$\prod_{j=1}^{|{\cal C}|} (- t^{|a|,{\cal C}}_{k,j})$ gives precisely
$(-1)^{|{\cal C}|}\; P^{{\cal C}}(t^{|a|}_k)\,R^{{\cal C}}(t^{|a|}_k)$
in (\ref{Sprp}). In the next level, each ${\cal C}_j$ in (\ref{Cj})
leads to a double cut on a path from some bottom vertex in
$t^{|a|,{\cal C}}_{k,j}$ to the root of $t^{|a|}_k$, and all double
cuts on paths from bottom to root of $t^{|a|}_k$ are obtained
(precisely once) in this way. The second cut is below the first one so
that the order of cuts is from top to bottom (and from left to right
anyway). By recursion one gets all possible cuts ${\cal C}_a$ of
$t^{|a|}_k$ contributing with the sign $(-1)^{|{\cal C}_a|}$ to the
antipode. \qed
\vskip 1ex

For $\delta^k_{ji,lm}$, the prescription (\ref{Sprp}) leads to the
following antipode:
\begin{eqnarray*}
S(\delta^k_{ji,lm}) &=& - 
\parbox{8mm}{\begin{picture}(6,18)
\put(1,15){$\bullet~^k_{ji}$}
\put(2,16){\line(0,-1){14}}
\put(1,8){$\bullet~_l$}
\put(1,1){$\bullet~_m$}
\end{picture}} - 
\parbox{14mm}{\begin{picture}(14,11)
\put(4.5,8){$\bullet~^k_{ji}$}
\put(5.5,9){\line(-1,-2){3.5}}
\put(5.5,9){\line(1,-2){3.5}}
\put(1,1){$\bullet~_l$}
\put(8,1){$\bullet~_m$} 
\end{picture}} \quad \longrightarrow \quad - \delta^k_{ji,lm}
\\
&+& \parbox{8mm}{\begin{picture}(6,18)
\put(1,15){$\bullet~^k_{ji}$}
\put(2,16){\line(0,-1){14}}
\put(0,4){---}
\put(1,8){$\bullet~_l$}
\put(1,1){$\bullet~_m$}
\end{picture}} \quad \longrightarrow \quad 
+ \delta^a_{lm} \delta^k_{ji,a}
\\
&+&\parbox{8mm}{\begin{picture}(6,18)
\put(1,15){$\bullet~^k_{ji}$}
\put(2,16){\line(0,-1){14}}
\put(0,11){---}
\put(1,8){$\bullet~_l$}
\put(1,1){$\bullet~_m$}
\end{picture}}\quad \longrightarrow \quad 
+ \delta^a_{jl,m} \delta^k_{ai}
+ \delta^a_{il,m} \delta^k_{ja}
- \delta^k_{al,m} \delta^a_{ji}
\\
&-&\parbox{8mm}{\begin{picture}(6,18)
\put(1,15){$\bullet~^k_{ji}$}
\put(2,16){\line(0,-1){14}}
\put(0,11){---}
\put(0,4){---}
\put(1,8){$\bullet~_l$}
\put(1,1){$\bullet~_m$}
\end{picture}}\quad \longrightarrow \quad 
\begin{array}[t]{l} - \delta^b_{jm} \delta^a_{bl} \delta^k_{ai}
- \delta^b_{lm} \delta^a_{jb} \delta^k_{ai}
+ \delta^a_{bm} \delta^b_{jl} \delta^k_{ai} \\
- \delta^b_{im} \delta^a_{bl} \delta^k_{ja}
- \delta^b_{lm} \delta^a_{ib} \delta^k_{ja}
+ \delta^a_{bm} \delta^b_{il} \delta^k_{ja} \\
+ \delta^b_{am} \delta^k_{bl} \delta^a_{ji}
+ \delta^b_{lm} \delta^k_{ab} \delta^a_{ji}
- \delta^k_{bm} \delta^b_{al} \delta^a_{ji}\end{array}
\\
&+& \parbox{14mm}{\begin{picture}(14,11)
\put(4.5,8){$\bullet~^k_{ji}$}
\put(5.5,9){\line(-1,-2){3.5}}
\put(5.5,9){\line(1,-2){3.5}}
\put(1,1){$\bullet~_l$}
\put(1.5,4){---}
\put(8,1){$\bullet~_m$} 
\end{picture}} \quad \longrightarrow \quad 
+ \delta^a_{jl} \delta^k_{ai,m}
+ \delta^a_{il} \delta^k_{ja,m}
- \delta^k_{al} \delta^a_{ji,m}
\\
&+& \parbox{14mm}{\begin{picture}(14,11)
\put(4.5,8){$\bullet~^k_{ji}$}
\put(5.5,9){\line(-1,-2){3.5}}
\put(5.5,9){\line(1,-2){3.5}}
\put(1,1){$\bullet~_l$}
\put(5.5,4){---}
\put(8,1){$\bullet~_m$} 
\end{picture}} \quad \longrightarrow \quad 
+ \delta^a_{jm} \delta^k_{ai,l}
+ \delta^a_{im} \delta^k_{ja,l}
- \delta^k_{am} \delta^a_{ji,l}
\\
&-& \parbox{14mm}{\begin{picture}(14,11)
\put(4.5,8){$\bullet~^k_{ji}$}
\put(5.5,9){\line(-1,-2){3.5}}
\put(5.5,9){\line(1,-2){3.5}}
\put(1,1){$\bullet~_l$}
\put(1.5,4){---}
\put(6,3){---}
\put(8,1){$\bullet~_m$} 
\end{picture}}  \quad \longrightarrow \quad 
\begin{array}[t]{l} - \delta^a_{jl} \delta^b_{am} \delta^k_{bi} 
- \delta^a_{jl} \delta^b_{im} \delta^k_{ab} 
+ \delta^a_{jl} \delta^k_{bm} \delta^b_{ai} 
\\
- \delta^a_{il} \delta^b_{jm}  \delta^k_{ba} 
- \delta^a_{il} \delta^b_{am}  \delta^k_{jb} 
+ \delta^a_{il} \delta^k_{bm}  \delta^b_{ja} 
\\[0.5ex]
+ \delta^k_{al} \delta^b_{jm} \delta^a_{bi} 
+ \delta^k_{al} \delta^b_{im}  \delta^a_{jb} 
- \delta^k_{al} \delta^a_{bm}  \delta^b_{ji} \end{array}
\end{eqnarray*}
One checks, using (\ref{Sd2}) and (\ref{lm})--(\ref{cut2}), the
antipode axioms (\ref{Sax}).

\section*{Acknowledgements}

First of all I would like to thank Alain Connes for his help on a
point were I was stuck. I profited very much from seminar notes by
Daniel Testard especially on the passage to the flat case. I hope to
include this in a future version. It is a pleasure to thank Bruno
Iochum, Thomas Krajewski, Serge Lazzarini, Thomas Sch\"ucker and
Daniel Testard for encouragement and numerous discussions. I am
grateful to Dirk Kreimer for a comment on eq.\ (\ref{rel}).

\end{document}